 \documentclass[final,5p,times,twocolumn]{elsarticle}

 \usepackage{graphicx}

\newcommand{\pT}{$p_T$ }
\newcommand{\mT} {$m_T$ }

\newcommand{\sNN}{$\sqrt{s_{_\mathrm{NN}}}$ }
\newcommand{\s}{$\sqrt{s}$ }
\newcommand{\pp}{$p$+$p$ }

\newcommand{\auau}{Au+Au }

\usepackage{amssymb}

\journal{Physics Letters B}

\begin{document}

\begin{frontmatter}



\title{Statistical Origin of Constituent-Quark Scaling in the QGP hadronization}


\author[addUSTC]{Zebo Tang}
\ead{zbtang@ustc.edu.cn}
\author[addUSTC,addPurdue]{Li Yi}
\author[addBNL]{Lijuan Ruan}
\author[addUSTC]{Ming Shao}
\author[addUSTC]{Hongfang Chen}
\author[addUSTC]{Cheng Li}
\author[addVECC]{Bedangadas Mohanty}
\author[addBNL]{Paul Sorensen}
\author[addBNL]{Aihong Tang}
\author[addBNL]{Zhangbu Xu}
\address[addUSTC]{University of Science \& Technology of China, Hefei 230026, China}
\address[addPurdue]{Purdue University, West Lafayette, Indiana 47907, USA}
\address[addBNL]{Brookhaven National Laboratory, Upton, New York 11973, USA}
\address[addVECC]{Variable Energy Cyclotron Centre, 1/AF, Bidhan Nagar, Kolkata 700064, India}

\begin{abstract}
Nonextensive statistics in a Blast-Wave model (TBW) is implemented
to describe the identified hadron production in relativistic p+p
and nucleus-nucleus collisions. Incorporating the core and corona
components within the TBW formalism allows us to describe
simultaneously some of the major observations in hadronic
observables at the Relativistic Heavy-Ion Collider (RHIC): the
Number of Constituent Quark Scaling (NCQ), the large radial and
elliptic flow, the effect of gluon saturation and the suppression
of hadron production at high transverse momentum ($p_T$) due to
jet quenching. In this formalism, the NCQ scaling at RHIC appears
as a consequence of non-equilibrium process. Our study also
provides concise reference distributions with a least $\chi^{2}$
fit of the available experimental data for future experiments and
models.
\end{abstract}


\end{frontmatter}


Several intriguing features were discovered in relativistic heavy
ion
collisions~\cite{starWhitePaper,phenixWhitePaper,Gyulassy:2004zy,muller:2004kk}
when particles  emerging from the Quark-Gluon Plasma were detected
by the experiments at RHIC. In Au+Au collisions, identified
particle yields integrated over the transverse momentum range
around the center-of-mass rapidity window have been shown to be at
equilibrium at the chemical freeze-out in a statistical
analysis~\cite{starWhitePaper,starPiKPLowPtPRL}. The hydrodynamic
model with proper equation of state and initial condition can
describe the anisotropic flow with small shear viscosity and
provides the notion of "perfect
liquid"~\cite{hydroReview,lacey:2006pn}. Furthermore, the
transverse momentum distributions of identified particles can be
described in a hydrodynamic-inspired model with a compact set of
parameters~\cite{starWhitePaper,starPiKPLowPtPRL,blastWave,Retiere_Lisa_PRC70,Tang:2008ud}.

However, in the intermediate $p_T$ range, particle production
exhibits grouping between baryons and mesons with baryons having
relatively higher yield and larger elliptic flow than the
mesons~\cite{starWhitePaper,Fries:2008hs}. This feature of
constituent quark scaling is not present in the hydrodynamics. A
microscopic quark coalescence at the hadronization seems to be
inescapable~\cite{muller:2004kk,Greco:2003xt}. At even higher
$p_T$, hard perturbative QCD processes (jets) are relevant.
Absorption of jets in the medium formed in A+A collisions has been
used for studying the properties of the
QGP~\cite{Gyulassy:2004zy,Zhang:2007ja,Liao:2009ni}. Even though
hydrodynamics with space-time evolution from an initial
condition~\cite{hydroReview} is so far the most realistic
simulation for bulk matter produced in relativistic heavy ion
collisions, its applicability is expected to breakdown for p+p and
peripheral A+A collisions at RHIC. Recent study showed that
hydrodynamics can not replace the microscopic hadronic cascade at
the late stage regardless of freeze-out and equation of state one
chooses~\cite{Song:2010aq} because the particle interactions may
be dominated by non-equilibrium hadronic
processes~\cite{Shao:2009mu}. In A+A collisions, the fluctuations
at initial impact due to Color-Glass Condensate (CGC) formation or
individual nucleon-nucleon collision may not be completely washed
out by subsequent interactions in either the QGP phase or hadronic
phase~\cite{Drescher:2000ec}. These effects leave footprints in
the spectra at low and intermediate $p_T$.

With its development and success of nonextensive statistics (also
known as Tsallis statistics~\cite{q-entropy}) in dealing with
non-equilibrated complex systems in condensed matter, many authors
have utilized Tsallis statistics to understand the particle
production in high-energy and nuclear
physics~\cite{Tang:2008ud,De:2007zza,Wilk:2008ue,Alberico:1999nh,Osada:2008sw,Biro:2003vz}.
Although the implications and understanding of the consequences of
such an application are still under investigation, the usual
Boltzmann distribution in an \mT exponential form can be readily
re-written as an \mT power-law (Levy) function~\cite{Wilk:1999dr}:
\begin{equation}
\frac{d^2N}{2\pi m_T dm_T
dy}\propto(1+\frac{q-1}{T}m_T)^{-1/(q-1)}\label{levy function}
\end{equation}
where the left-hand side is the invariant differential particle
yield and $q$ is a parameter characterizing the degree of
non-equilibrium, $m_T=\sqrt{m^{2}+p_T^2}$ is the transverse mass
of the given particle with mass of $m$ and $T$ is related to the
temperature of the system. The distribution can be derived from
the usual procedure in statistical mechanics, starting from a
non-equilibrium $q$-entropy~\cite{q-entropy}.
The successful application of Levy functions ($TBW_{pp}$) to the
spectra in p+p collisions at RHIC resulted in $q$ values
significantly larger than unity and are different between the
groups of baryons and mesons~\cite{Tang:2008ud,Adare:2010fe}.
However, in the central Au+Au collisions, the spectra at low $p_T$
show characteristic Boltzmann distribution with $q$ value being
close to unity ($q\rightarrow1$, Eq.~\ref{levy function} becomes a
Boltzmann distribution) even though there are still significant
power-law tails with considerable particle yields at high
$p_T$~\cite{Tang:2008ud}. In addition to the escaping jets at high
$p_T$, coalescence with non-equilibrated quarks has also been
proposed to study the power-law behavior~\cite{Biro:2008hz}.
Difficulty in accounting for these processes so far seems to be a
major limitation of the TBW statistical description of the
experimental data over a wide $p_T$ range~\cite{Tang:2008ud}. To
bridge the hydrodynamic nature of the spectra at low $p_T$ and
power-law tails at high $p_T$ with smooth transition at
intermediate $p_T$, models which include a hot and dense core with
a corona of jet-like process have been
proposed~\cite{Liao:2009ni,Werner:2007bf}.

\begin{figure}[htp]
\includegraphics[width=0.44\textwidth]{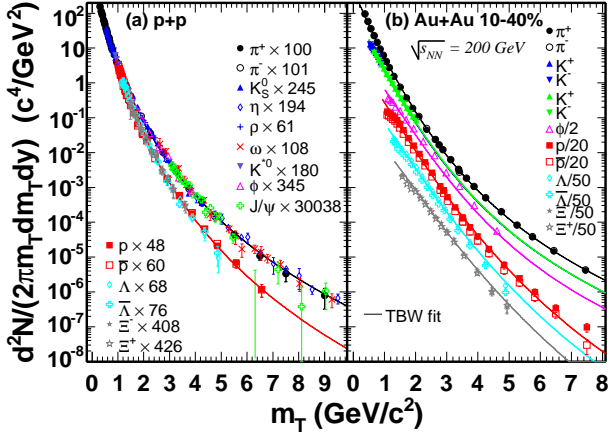}
\caption{(Color Online) Identified particle transverse mass
spectra in \pp collisions (a) and 10-40\% \auau collisions (b) at
\sNN = 200 GeV. The symbols represent experimental data points.
The curves represent the TBW fit. Only fits to the particles are
shown since the model has the same spectral shapes for particles
and anti-particles. For plotting in panel (a), the spectra of
meson (baryon) are scaled to match that of $\pi^{+}$ ($p$) at
$m_T=1.5$ GeV/$c^2$ for $\phi$ ($\Xi^\pm$), at 4 GeV/$c^2$ for
$J/\psi$ and at 1 GeV/$c^2$ for the rest.} \label{fig:spectra}
\end{figure}

In this paper, we present the procedure of implementing
nonextensive statistics in the Blast-Wave model (TBW) with
azimuthally anisotropic particle emission, and use it to fit the
identified particle spectra and for the first time to elliptic
flow at mid-rapidity at RHIC. The model uses the TBW function
obtained from p+p data~\cite{Tang:2008ud} as corona and an
additional TBW function as core to fit Au+Au data. The formalism
thus provides a systematic comparison between p+p and central A+A
collisions in one macroscopic statistical model framework and
gives an accurate numeric description of the experimental data
over a wide range of $p_T$. Examples of such successful
applications in the related subjects are the chemical fit to the
particle yields~\cite{Andronic:2003zv,Cleymans:2008mt} and the
global fit of the parton distribution function (PDF) of
proton~\cite{Lai:1994bb}. Good TBW fits can also offer a simple
formula for developing ideas and building models in a reasonably
realistic
environment~\cite{starPiKPLowPtPRL,blastWave,Retiere_Lisa_PRC70,Liao:2009ni,Zhao:2007hh,Gavin:2008ev,Shuryak:2007fu,fuqiangLongPID},
and provide a practical experimental tool to extract particle
yields by extrapolating to unmeasured kinematic ranges.


\begin{figure}[ht]
\includegraphics[width=0.43\textwidth]{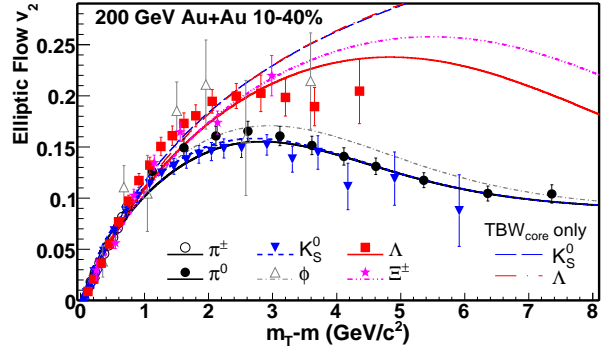}
\caption{(Color Online) Identified particle $v_2$ in 10-40\% \auau
collisions. The x-axis is depicted by the kinetic energy
($m_T-m$)~\cite{lacey:2006pn}, showing the scaling of $v_2$ at low
$p_T$, and grouping of baryons and mesons at the intermediate
$p_T$ range. The curves represent the TBW fit
(Eq.~\ref{eq:core-corona}) and the characteristic NCQ scaling.
Also shown are the  indistinguishable $K^0_S$ and $\Lambda$ curves
from $TBW_{core}$ alone over the entire range.} \label{fig:v2Fit}
\end{figure}
To take into account collective flow and azimuthal anisotropy in
the transverse direction in relativistic heavy ion collisions,
Levy distribution needs to be embedded in the framework of
hydrodynamic expansion~\cite{Wilk:2008ue}. We follow the recipe of
the Blast-Wave model provided in
literature~\cite{starPiKPLowPtPRL,blastWave,Retiere_Lisa_PRC70,fuqiangLongPID},
and change sources of particle emission from a Boltzmann
distribution to a Levy distribution~\cite{Tang:2008ud}:
\begin{eqnarray}
\frac{dN}{m_T dm_Td\phi} \propto m_T
\int_{0}^{2\pi}d\phi_s\int_{-y_b}^{+y_b}
dy~e^{\sqrt{y_b^{2}-y^{2}}}\cosh(y) \nonumber\\
\times \int_{0}^{R}rdr(1+\frac{q-1}{T}E_{T})^{-1/(q-1)},
\label{tbw2}
\end{eqnarray}
where $y_b=\ln{(\sqrt{s_{_{NN}}}/m_{_{N}})}$~\cite{Wong:2008ex} is
the beam rapidity and the rapidity distribution can be
approximated as a Gaussian with a width of $\sigma_y=2.27\pm0.02$
at the center-of-mass energy of $\sqrt{s_{_{NN}}}=200$
GeV~\cite{Bearden:2004yx,Pawan:PhysRevC71}, transverse energy
$E_{T}=m_T\cosh(y)\cosh(\rho)-p_T\sinh(\rho)\cos(\phi_b-\phi)$,
 $\rho =\sqrt{(r\cos{(\phi_s)}/R_X)^{2}+(r\sin{(\phi_s)}/R_Y)^{2}}$$(\rho_{0}+\rho_{2}\cos{(2\phi_{b})})$
is the flow profile in transverse rapidity, and
$\tan{(\phi_b)}=(R_X/R_Y)^{2}\tan{(\phi_s)}$ relates the azimuthal
angle of the coordinate space ($\phi_s$) to the angle of the flow
direction ($\phi_b$) of the emitting
source~\cite{Retiere_Lisa_PRC70}. Equation~\ref{tbw2} extends the
nonextensive statistics in a blast-wave model~\cite{Tang:2008ud}
to incorporate particle emission from an elliptic source ($R_X$
and $R_Y$ are the axes in the coordinate space) with an elliptic
expansion ($\rho_{0}$ and $\rho_{2}$)~\cite{Retiere_Lisa_PRC70}.
In addition to this core component, it is important to include the
corona with jet-like particle emission at high momentum resembling
an ensemble of individual p+p
collisions~\cite{Liao:2009ni,Werner:2007bf}. The combined core and
corona formula reads:
\begin{eqnarray}
\frac{dN}{m_T dm_Td\phi}|_{AA} = TBW_{core} \nonumber\\
+f_{pp}N_{bin}\epsilon(1+v_{2}^{jet}\cos{(2\phi)})TBW_{pp},
\label{eq:core-corona}
\end{eqnarray}
where $TBW_{core}$ is from Eq.~\ref{tbw2}, $f_{pp}$ and
$v_{2}^{jet}$ represent the fraction and the anisotropy of the
escaping jet comparing to the expected number of binary p+p
collisions in Au+Au collisions ($N_{bin}$),
$\epsilon=p_{T}^{2}/(p_{T}^{2}+Q_{S}^{2})$ takes into account the
gluon saturation effect from p+p to Au+Au collisions with
saturation scale $Q_S=1.5$ GeV/$c$~\cite{SchaffnerBielich:2001qj},
and $TBW_{pp}$ are the adopted fit results from the spectra in p+p
collisions without any additional free parameter for either
baryons or mesons. Since radial flow velocity $\rho$ was found to
be zero in \pp collisions~\cite{Tang:2008ud}, $TBW_{pp}$ can be
simplified as:
\begin{eqnarray}
TBW_{pp} \propto m_T \int_{-y_b}^{+y_b}
dy~e^{\sqrt{y_b^{2}-y^{2}}}\cosh(y) \nonumber\\
\times (1+\frac{q-1}{T} m_T \cosh(y))^{-1/(q-1)}, \label{tbwpp}
\end{eqnarray}

\begin{figure}[hb]
\includegraphics[width=0.43\textwidth]{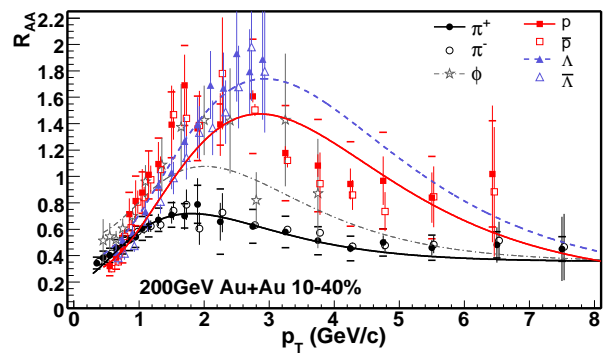}
\caption{(Color Online) Comparison of TBW fit (lines) with STAR
measurements of $R_{AA}$ in 10-40\% \auau collisions. The syst.
errors (horizontal bars) and stat. errors (vertical lines) are
shown separately for protons and charged pions. Systematic and
statistical errors are added quadratically in the fit, and are
shown as such in the plot except the protons and charged pions.}
\label{fig:raa}
\end{figure}

The STAR and PHENIX collaborations have published the most
complete series of particle spectra and $v_2$ at mid-rapidity for
p+p and Au+Au collisions at \sNN = 200 GeV. The identified
particle spectra and $v_2$ include $\pi^{\pm}$, $K^{\pm}$, $K_S$,
$K^*$, $p$, $\phi$, $\Lambda$, $\Xi^-$, $\bar{p}$,
$\bar{\Lambda}$, and $\Xi^+$ in STAR
publications~\cite{starWhitePaper,starPiKPLowPtPRL,starKstar,starPhiAuAu200,ppSplit,
Abelev:2006jr,Abelev:2007ra,Adams:2003qm,Adams:2004ux,
Adams:2006ke,Adams:2006nd,Adams:2004bi,Abelev:2008ed,rho:2003cc}.
The $\eta$, $\eta'$ and $\omega$ spectra in \pp collisions,
$K^{\pm}$ spectra and $\pi^0$ $v_2$ in \auau collisions are from
PHENIX
publications~\cite{PHENIXetapp,PHENIXomegapp,PHENIXJpsipp,PHENIXKaon,Adare:2010sp,PHENIXppNeutral}.
Figure~\ref{fig:spectra}.a shows the invariant differential yields
in \pp collisions at \s = 200 GeV. All of the mesons or baryons
have the common $m_T$ spectra shape. The solid lines represent the
fit to Eq.~\ref{tbwpp} for mesons and baryons separately. The
common fit parameters and best $\chi^2$ per fitting degree of
freedom (nDoF) are listed in Tab.~\ref{tablepp}. In addition to
the common fit parameters, a global normalization factor was
applied for each particle spectrum. The $p_T$-integrated cross
section $d\sigma/dy$ for each particle species obtained from the
fit are listed in Tab.~\ref{tab:dndy}. Figure~\ref{fig:spectra}.b
shows the invariant differential yields together with our fit
results in 10-40\% centrality Au+Au collisions. The results of
$v_2$ from the simultaneous fit with spectra are displayed in
Fig.~\ref{fig:v2Fit}. The fit parameters and best $\chi^2$ per
nDoF are tabulated in Tab.~\ref{table}. In addition to these
parameters common to all particles, a fit parameter is required as
a normalization factor for $TBW_{core}$ of each particle spectrum.
The $TBW_{pp}$ in the fit function is directly adopted from the
fit results shown in Fig.~\ref{fig:spectra}.a for each particle
spices without any additional free parameter. The $R_{AA}$ (ratio
of the $N_{bin}$ normalized $p_T$ spectra in A+A collisions to the
underlying p+p spectrum) from model reproduces the data very well,
as shown in Fig.~\ref{fig:raa}.

\begin{table}
\caption{Values of parameters and best $\chi^{2}$ from TBW fit of
Eq.~\ref{tbwpp} to identified particle \pT spectra in \pp
collisions at RHIC. The uncorrelated systematic errors are
included in the fit. \label{tablepp}}
\begin{tabular}{c|ccccccc}
\hline
&  T (MeV) & $q-1$ &$\chi^2$/nDoF\\
\hline Mesons&89.9 $\pm$ 0.7 & 0.0955 $\pm$ 0.0006 & 283/257\\
Baryons&68.5 $\pm$ 4.0 & 0.0855 $\pm$ 0.0019 & 151/128\\
\hline
\end{tabular}
\end{table}

\begin{table*}[htb]
\caption{Cross sections $d\sigma/dy$ in mb of different particles
at mid-rapidity in \pp collisions at \sNN = 200~GeV. ``$TBW_{pp}$"
column show the values obtained from fit results shown in
Fig.~\ref{fig:spectra}.a. ``Data Ref." column shows where are the
data points from. ``PHENIX" column shows the values obtained by
fits to PHENIX measured spectra with the Tsallis functional form
as described in Ref.~\cite{PHENIXppNeutral}. "Published" column
shows the values published by experiments. An additional 9.7\%
systematic uncertainty should be added to all $d\sigma/dy$ values
listed in the column ``PHENIX'' and 12\% to the values in the
column ``'STAR' to account for the trigger uncertainties. Values
in the column ``Published'' are also given without these
systematic uncertainties. The column ``S.M.'' is the prediction of
the statistical model~\cite{StatModel}. All errors are the
combined statistical and systematic
uncertainties.}\label{tab:dndy}
\begin{tabular}{lccccccc}
\hline
Particle                              & $TBW_{pp}$    & Data Ref.             & PHENIX~\cite{PHENIXppNeutral}     & STAR          & S.M.~\cite{StatModel}                   \\
\hline
$\pi^{0}$                             &               &                       &$41.4\pm5.8$    &               & 46.9   \\
$\pi^{+}$                             & 42.4          & \cite{starPiKPLowPtPRL,Adams:2006nd}   &$39.4\pm7.3$    &$43.2\pm3.3$~\cite{fuqiangLongPID}   & 42.1          \\
$\pi^{-}$                             & 41.9          & \cite{starPiKPLowPtPRL,Adams:2006nd}   &$38.6\pm7.2$    &$42.6\pm3.3$~\cite{fuqiangLongPID}   & 41.5       \\
$K^{+}$                               & 4.69          & \cite{starPiKPLowPtPRL,Adams:2003qm,ppSplit} &$4.57\pm0.61$    &$4.50\pm0.39$ \cite{fuqiangLongPID}  & 4.57  \\
$K^{-}$                               & 4.57          & \cite{starPiKPLowPtPRL,Adams:2003qm,ppSplit} &$4.20\pm0.51$   &$4.35\pm0.39$ \cite{fuqiangLongPID}  & 4.38  \\
$K^{0}_{S}$                           & 4.73          & \cite{ppSplit}        &$5.28\pm0.53$   &$4.02\pm0.34$ \cite{ppSplit}   & 4.40  \\
$\eta$                                & 4.94          & \cite{PHENIXetapp}    &$4.47\pm0.96$   &               & 4.93      \\
$\rho$                                & 6.79          & \cite{rho:2003cc}     &                &$7.8\pm1.2$ \cite{rho:2003cc}    & 5.58             \\
$\omega$                              & 3.75          & \cite{PHENIXomegapp}  &$3.65\pm0.77$   & & 5.03                  \\
$\eta'$                               & 0.58          & \cite{PHENIXppNeutral}&$0.62\pm0.17$   &               & 0.365              \\
$(K^{*+}+K^{*-})/2$                   & 1.51          & \cite{starKstar}      &               & & 1.57      &    \\
$(K^{*0}+\bar{K}^{*0})/2$             & 1.54          & \cite{starKstar}      &                &$1.52\pm0.19$ \cite{starKstar}  & 1.55          \\
$\phi$                                & 0.521         & \cite{Adams:2004ux}   &$0.421\pm0.055$ &$0.540\pm0.086$ \cite{Adams:2004ux}& 0.339                  \\
$J/\psi$ ($\times10^{3}$)             & 0.741         & \cite{PHENIXJpsipp}   &$0.759\pm0.053$ &  &                   \\
$\psi'$  ($\times10^{3}$)             &               &                       &$0.133\pm0.031$ &&           \\
\hline
$p$                                   & 3.96          & \cite{starPiKPLowPtPRL,Adams:2006nd}   &                &$4.14\pm0.30$ \cite{fuqiangLongPID}   & 4.47        \\
$\bar{p}$                             & 3.19          & \cite{starPiKPLowPtPRL,Adams:2006nd}   &                &$3.39\pm0.36$ \cite{fuqiangLongPID}  & 3.59        \\
$\Lambda$                             & 1.36          & \cite{ppSplit}        &                &$1.31\pm0.12$ \cite{ppSplit}   & 1.30         \\
$\bar{\Lambda}$                       & 1.21          & \cite{ppSplit}        &                &$1.19\pm0.11$ \cite{ppSplit}   & 1.11      \\
$\Xi^{-}$                             & 0.098         & \cite{ppSplit}        &                &$0.078\pm0.028$ \cite{ppSplit} & 0.092     \\
$\bar{\Xi}^{+}$                       & 0.094         & \cite{ppSplit}        &                &$0.087\pm0.031$ \cite{ppSplit} & 0.082      \\
$\Sigma^{*+}+\Sigma^{*-}$             & 0.359         & \cite{extrotic_hyperon} &                &$0.321\pm0.044$ \cite{extrotic_hyperon}& 0.308     \\
$\bar{\Sigma}^{*+}+\bar{\Sigma}^{*-}$ & 0.313         & \cite{extrotic_hyperon} &                &$0.267\pm0.038$ \cite{extrotic_hyperon}& 0.260     \\
$\bar{\Lambda}^{*}+\Lambda^{*}$       & 0.125         & \cite{extrotic_hyperon}&                &$0.104\pm0.017$ \cite{extrotic_hyperon} & 0.168     \\
$\Omega^{-}+\bar{\Omega}^{+}$ ($\times10^{3}$)& 12.3  & \cite{ppSplit}         &                &$10.2\pm5.7$ \cite{ppSplit}   & 17.1   \\
\hline
\end{tabular}
\end{table*}

\begin{table*}
\caption{Values of parameters and best $\chi^{2}$ from TBW fit of
Eq.~\ref{eq:core-corona} to identified particle \pT spectra and
$v_2$ in 10-40\% Au+Au collisions at RHIC.
The uncorrelated systematic errors are included in the fit. Syst.
error of STAR $v_2\{EP\}$ are taken to be
$15\%/\sqrt{12}$~\cite{Adams:2004bi}. The spectra contribute
359/244 to the $\chi^{2}$/nDoF. $Q_S=2.1\pm0.2$ GeV/$c$ with
$\chi^{2}$/nDoF=497/295 when $Q_S$ is a free parameter. Including
the $v_{2}\{EP\}$ of $K^{\pm}$ and $\bar{p}$ at low
$p_T$~\cite{YTBai} increases the best $\chi^{2}$/nDoF to 769/307.
\label{table}}
\begin{tabular}{cccccccc}
\hline
$\rho_{0}$&$\rho_{2}$&$R_X/R_Y$&T (MeV)&$q-1$&$f_{pp}$&$v_2^{jet}$&$\chi^2$/nDoF\\
\hline $0.654 \pm 0.002$ & $0.199 \pm 0.002$ & $0.871 \pm 0.004$ & $128 \pm 2$ & $0.044 \pm 0.001$ & $0.36 \pm 0.01$ & ($8.7 \pm 0.4$) \% &506/296\\
\hline
\end{tabular}
\end{table*}

The Blast-Wave model with nonextensive statistics and azimuthal
anisotropy and a core-corona composition has allowed high quality
fits to spectra and elliptic flow over a broad $p_T$ range. The
striking feature of the experimental observations related to the
NCQ scaling is well reproduced. The results can be summarized as
follows:
1) The bulk core alone ($TBW_{core}$) with finite $q$ value can
fit the data very well at low
$p_T$~\cite{Tang:2008ud,Shao:2009mu}. The new extension to $v_2$
continues to provide high quality fits. The system produces
maximum radial flow velocities of
$\tanh{(\rho_{0}+\rho_{2})}=0.69c$ and
$\tanh{(\rho_{0}-\rho_{2})}=0.43c$ along x-axis (the reaction
plane) and y-axis, respectively. This demonstrates that the bulk
system can be described with a few macroscopic parameters and is
qualified as a thermodynamic state.
2) Figure~\ref{fig:raa} shows that the $R_{AA}$ of the
experimental data points at low $p_T$ are below those at high
$p_T$. The $\epsilon$ parameter necessarily brings the p+p
component at low $p_T$ down to a subdominant fraction. This
necessity may be partly due to the modification of the jet
low-$p_T$ component by the bulk~\cite{Liao:2009ni}.
3) The non-equilibrated component in the bulk core produces a
power-law tail in spectra and high $v_2$ at the intermediate
$p_T$.
4) The baryon and meson yields ($TBW_{pp}$) are grouped in p+p
collisions with baryon yields systematically lower than meson
yields as observed in the experimental
data~\cite{Tang:2008ud,Werner:2007bf,ppSplit}.
5) The combination of non-equilibrium tail from core and the
baryon-meson separation from corona brings down the bulk $v_2$,
produces the baryon enhancement and the NCQ scaling at the
intermediate $p_T$.
6) The medium quenches the jet and reduces it to a fraction
($f_{pp}=0.36$) of its underlying binary nucleon-nucleon
collisions, resulting in a finite azimuthally anisotropic emission
($v_{2}^{jet}=8.7\%$)~\cite{Liao:2008dk}.
7) The high-precision ($\pm10^{-4}$ stat.) experimental data
points concentrate at low $p_T$, dominating the fit $\chi^{2}$.
Additional high-quality data in higher $p_T$ range (e.g. $v_2$ of
baryons) will balance the contributions to the $\chi^{2}$ from
components with different physical origins. The not-quite-ideal
$\chi^{2}$/nDoF value indicates significant tensions among the
different datasets and model, and warrants further detailed
assessment and categorization similar to the PDF
fit~\cite{Lai:PhysRevD82} in the future.


This TBW model describes the number of constituent quark (NCQ)
dependencies observed in the data which have also been described
in recombination
models~\cite{muller:2004kk,Greco:2003xt,Biro:2008hz,Hwa:2004ng}.
The NCQ dependence indicates that baryon production at
intermediate $p_T$ increases faster going from p+p to Au+Au
collisions than meson production does. The division into a core
and corona component in Eq.~\ref{eq:core-corona} represents a
division into two samples: a core and corona where the core has a
larger fraction of baryons than the corona. We describe both the
core and corona with a TBW statistical model. But whereas the
recombination picture attempts to provide a microscopic
description for why the bulk in Au+Au collisions has a larger
fraction of baryons, the macroscopic TBW statistical model only
describes the existing states as they are, and is not sensitive to
the underlying mechanism responsible for producing those states.
Regardless of whether the $TBW_{pp}$ component is indeed from
individual p+p collision or part of the hadrons from coalescence
of the escaping partons, Eq.~\ref{eq:core-corona} does provide a
unified tool to describe simultaneously a variety of measurements
over a wide range in $p_T$ for future studies. These results also
provide a macroscopic foundation for discussing the entropy issues
associated with the system's underlying microscopic
subprocesses~\cite{Biro:2003vz,Biro:2008hz,He:2010vw,Maroney:PhysRevE80}.

In summary, we have implemented the nonextensive statistics in a
Blast-Wave model and incorporated a core-corona model to describe
a complete data set on identified particle spectra and elliptic
flow versus transverse momenta at mid-rapidity measured at RHIC.
Such a formalism simultaneously describes several novel
observations reported at RHIC over a broad $p_T$ range.
Specifically it provides an alternative physical picture, through
the role of the core-corona and non-equilibrium effects in
understanding the baryon-meson differences in the nuclear
modification factor and the elliptic flow at intermediate $p_T$.
The macroscopic nonextensive statistics provides a complementary
microscopic method to the hydrodynamics and hadronic cascade for
studying the evolution of the nucleus-nucleus collisions and the
properties of QGP.

\section*{Acknowledgments}
The authors thank Dr. Yuting Bai for clarification of STAR
$v_2\{2\}$ and $v_2\{EP\}$ data, and Drs. B. Christie, J. Dunlop,
J. Liao, V. Koch, Profs. F.Q. Wang, Q. Wang and G. Wilk for
discussions. This work was supported in part by the Offices of NP
and HEP within the U.S. DOE Office of Science under the contracts
of DE-FG02-88ER40412 and DE-AC02-98CH10886. USTC group is
supported in part by NNSF of China under Grant No. 11005103 and
10835005. BM is supported by DAE-BRNS project Sanction No.
2010/21/15-BRNS/2026.





\bibliographystyle{elsarticle-num}
\bibliography{TsallisBlastWave_EllipticFlow_v15}







\end{document}